\documentclass[prd,showpacs]{revtex4}
\usepackage{amsmath,amssymb}
\usepackage{xcolor}

\numberwithin{equation}{section}
\begin{document}
\title{Casimir effect of a doubly Lorentz-violating scalar in magnetic field}
\author{Andrea Erdas}
\email{aerdas@loyola.edu}
\affiliation{Department of Physics, Loyola University Maryland, 4501 North Charles Street,
Baltimore, Maryland 21210, USA}
\date{October 4, 2024}
\begin {abstract} 
In this work I study the Casimir effect caused by a charged and massive scalar field that violates Lorentz symmetry in an aether-like and CPT even manner, by direct coupling between the field derivatives and two fixed orthogonal unit vectors. I call this a "double" Lorentz violation.  The field will satisfy either Dirichlet or mixed boundary conditions on a pair of plane parallel plates. I use the generalized zeta function technique to examine the effect of a uniform magnetic field, perpendicular to the plates, on the Casimir energy and pressure. Three different pairs of mutual directions of the unit vectors are possible: timelike and spacelike perpendicular to the magnetic field, timelike and spacelike parallel to the magnetic field, spacelike perpendicular and spacelike parallel to the magnetic field. I fully examine all of them for both types of boundary conditions listed above and, in all cases and for both types of boundary conditions, I obtain simple expressions of the zeta function, Casimir energy and pressure in the three asymptotic limits of strong magnetic field, large mass, and small plate distance.
\end {abstract}
\maketitle
%%%%%%%%%%%%%%%%%%%%%%%%%%%%%%%%%%%%%%%%%%%%%%%%%%%%%%%%%%%%%
\section{ Introduction}
\label{1}
The Casimir effect was first presented as a theoretical prediction in 1948 \cite{Casimir:1948dh}. In its simplest form, it argues that two uncharged conducting plates facing each other in vacuum feel an attractive force due to quantum field theory effects. Experimental evidence for it \cite{Sparnaay:1958wg} started to arrive ten years after Casimir first published his work, and many and progressively more accurate confirming experiments have been published since \cite{Bordag:2001qi,Bordag:2009zz}. For this effect to manifest itself in its simplest form, we need to put two parallel plates in a region of space filled with one or more quantum fields. The force between the plates strongly depends on the boundary conditions of the quantum field(s) at the plates: for Dirichlet boundary conditions at both plates, the force is attractive, the same for Neumann boundary conditions. However, if boundary conditions are mixed, i.e. Dirichlet on one plate and Neumann on the other, the Casimir force is repulsive \cite{Boyer:1974}.

The Lorentz symmetry is one of the cornerstones of standard quantum field theory. However, more recent theories propose scenarios where a violation of Lorentz symmetry is possible and leads to an anisotropy of spacetime that could be due to a string theory mechanism that produces non-vanishing vacuum expectation values of some vector  or tensor components of a quantum field \cite{Kostelecky:1988zi}, or to spacetime non-commutativity \cite{Carroll:2001ws}, or to a varying coupling constant \cite{Horava:2009uw}.
Many attempts to search for a theory of quantum gravity, or within string theory, have produced spacetime anisotropy as a consequence \cite{Anchordoqui:2003ij,Bertolami:1997iy,Kostelecky:2002ca}. While the Lorentz symmetry breaking happens at the Planck scale, the effects of 
the spacetime anisotropy are observable at low energy through the Casimir effect, whether the effect is due to a scalar field or a fermion field \cite{Frank:2006ww,Ulion:2015kjx,Cruz:2017kfo,Cruz:2018bqt,daSilva:2019iwn,Droguett:2024tpe,Borquez:2023ajx,Escobar:2020pes,Escobar:2020gvb,Araujo:2023iyi}. Some of these papers \cite{Cruz:2017kfo,Cruz:2018bqt} propose and investigate a scalar field that satisfies a modified Klein-Gordon equation, a model that breaks Lorentz invariance in a CPT-even, aether-like manner through the direct coupling between the field derivatives and a fixed, arbitrary four-vector. A study was done to analyze the effect on the Casimir effect of the topological mass of a self-interacting scalar field that violates Lorentz symmetry in an aether-like manner \cite{Cruz:2020zkc}. A recent paper \cite{deMello:2022tuv}  expanded this line of research by carrying out the first detailed investigation where the Lorentz symmetry is broken in such a way that two distinct and orthogonal preferred directions appear in four-dimensional spacetime. This paper continues in this line of work by presenting a more complex model where a scalar field breaks Lorentz invariance in an aether-like and CPT even manner, by way of a direct coupling between the field derivatives and two fixed orthogonal unit vectors.  This could be a first model, for example, for a nontrivial biaxial vacuum for QED. While one could, in principle, consider a broken Lorentz symmetry where more than two preferred directions appear in an aether-like and CPT even manner, such a study has not appeared in the literature and is beyond the scope of this paper.

Magnetic corrections to the Casimir effect in Lorentz symmetric spacetime have been investigated extensively in vacuum and in a medium at finite temperature. See, for example, \cite{Cougo-Pinto:1998jun,Cougo-Pinto:1998fpo,Erdas:2013jga,Erdas:2013dha}. Some recent papers examined the Casimir effect when the Lorentz symmetry is broken and a magnetic field is present \cite{Erdas:2020ilo,Erdas:2021xvv,Rohim:2023tmy,Santos:2022tbq}, with interesting findings about the way Lorentz asymmetry and the presence of a magnetic field modify the Casimir energy and force. This paper intends to extend this line of research and produce theoretical predictions of the magnetic effects on the model of a charged scalar field that breaks Lorentz invariance in an aether-like and CPT even manner, by direct coupling between the field derivatives and two fixed orthogonal unit vectors. I will investigate a charged scalar field that violates Lorentz symmetry as described by \cite{deMello:2022tuv} and satisfies Dirichlet or mixed boundary conditions on a pair of parallel plates at distance $a$ from each other, with a uniform magnetic field $\bf B$ present and perpendicular to the plates. I will examine all possible combinations of the two orthogonal unit vectors, $u^\mu$ and $v^\mu$, and calculate Casimir energy and pressure for all situations. Most Casimir energy calculations require a regularization method to obtain the correct finite result. In this work I will use the generalized zeta function technique  \cite{Hawking:1976ja}, while the authors of Ref. \cite{deMello:2022tuv} used the Abel-Plana method.

Sec. \ref{2} of this paper briefly describes the model of a charged scalar field that "doubly" breaks Lorentz invariance, identifies the three direction possibilities for $u^\mu$ and $v^\mu$, and calculates the zeta function for each of these situations. In Sec. \ref{3} I examine the case where $u^\mu$ is timelike and $v^\mu$ is spacelike and orthogonal to the magnetic field. I evaluate the zeta function under three different approximations, strong magnetic field, large mass, and small plate distance and obtain simple analytic expressions of the Casimir energy and pressure in each of these three approximations. In Sec. \ref{4} I examine the case of a timelike $u$ and a spacelike $v$ that is parallel to the magnetic field, and evaluate Casimir energy and pressure under the three approximations listed above. In Sec.  \ref{5} I consider the case where $u$ and $v$ are spacelike and one is perpendicular and the other parallel to the magnetic field, and obtain Casimir energy and pressure in the three approximations discussed. My conclusions and a discussion of my results are in Sec. \ref{6}. In the Appendix I show the details of the lengthier calculations of zeta function and Casimir energy in the small plate distance approximation.
%%%%%%%%%%%%%%%%%%%%%%%%%%%%%%%%%%%%%%%%%%%%%%%%%%%%%%%%%%%%%%%%%%%%%
\section{The model and the zeta functions}
\label{2}

In this paper I take $\hbar=c=1$ and will investigate the Casimir effect due to a complex scalar field $\phi$ of mass $m$ and charge $e$ that violates Lorentz symmetry in two orthogonal directions. I use the theoretical model proposed by Bezerra de Mello and Cruz in Ref. \cite{deMello:2022tuv}, where a real scalar field breaks Lorentz symmetry in an aether-like and CPT even manner, by way of a direct coupling between the field derivatives and two fixed orthogonal unit vectors $u^\mu$ and $v^\mu$. This model's Lagrangian for the real scalar field $\phi$ is
\begin{equation}
{\cal L}= {1\over 2}\left[\left(\partial^\mu\phi\right)\partial_\mu\phi+\lambda_1\left(u\cdot\partial \phi\right)^2+\lambda_2\left(v\cdot\partial \phi\right)^2-m^2\phi^2\right],
\label{L}
\end{equation}
where $\lambda_1\ll1$ and $\lambda_2\ll 1$ are dimensionless parameters that determine the strength of the Lorentz violation in the $u$ and $v$ direction respectively. In Ref. \cite{deMello:2022tuv} the authors use $\cal L$ to derive the modified Klein-Gordon equation for $\phi$
\begin{equation}
\left[ \Box +\lambda_1\left(u\cdot\partial \right)^2+\lambda_2\left(v\cdot\partial \right)^2+m^2\right]\phi(x)=0.
\label{KG}
\end{equation}
The generalization of Eqs. (\ref{L}) and (\ref{KG}) to a complex scalar field is completely straightforward. I intend to study how this anisotropy of space-time and the presence of a constant magnetic field perpendicular to the plates modify the Casimir effect. I consider two identical square plates of side $L$, perpendicular to the $z$-axis and located one at $z=0$ and one at $z=a$, with $a\ll L$ being the plate distance. A constant magnetic field $\bf B$ pointing in the $z$-direction will be introduced shortly. The main tool used in this theoretical study is the generalized zeta function regularization method \cite{Hawking:1976ja}, and I will investigate the case of Dirichlet and mixed (Dirichlet-Neumann) boundary conditions at the plates for the scalar field $\phi$. The case of Neumann boundary conditions is straightforward and produces the same results as that of Dirichlet boundary conditions. I will study all three possible combinations of directions of the orthogonal unit vectors $u^\mu$ and $v^\mu$,

{\bf \textcircled{1}} $u$ is timelike and $v$ is spacelike and perpendicular ($\perp$) to $\bf B$: $u^\mu = (1,0,0,0)$, and $v^\mu = (0,{1\over {\sqrt 2}},{1\over {\sqrt 2}},0)$,

{\bf \textcircled{2}} $u$ is timelike and $v$ is spacelike and parallel ($\parallel$)to $\bf B$: $u^\mu = (1,0,0,0)$, and $v^\mu = (0,0,0,1)$,

{\bf \textcircled{3}} $u$ is spacelike and $\perp$ to $\bf B$ and $v$ is spacelike and $\parallel$ to $\bf B$: $u^\mu = (0,{1\over {\sqrt 2}},{1\over {\sqrt 2}},,0)$, and $v^\mu = (0,0,0,1)$.

\noindent
I begin by switching to Euclidean time $\tau$ and writing the eigenvalues of the Klein Gordon operator $D_E$ of Eq. (\ref{KG})
\begin{equation}
D_E=-{\partial^2\over \partial \tau^2}-\nabla^2+\lambda_1\left(u\cdot\partial \right)^2+\lambda_2\left(v\cdot\partial \right)^2+m^2,
\label{KGoperator}
\end{equation}
notice that at this stage I have not yet introduced the magnetic field. I will now derive and list the eigenvalues of $D_E$ for all possible combinations of directions of $u$ and $v$ and of Dirichlet and mixed boundary conditions.
\begin{itemize}
\item Case {\bf \textcircled{1}}: $u$ is timelike and $v$ is $\perp$ to $\bf B$.

When $u$ is timelike
\begin{equation}
-{\partial^2\over \partial \tau^2}+\lambda_1\left(u\cdot\partial \right)^2=-(1+\lambda_1){\partial^2\over \partial \tau^2}.
\label{eigenvalues01}
\end{equation}
The eigenfunctions of this differential operator are plane waves with corresponding eigenvalues
\begin{equation}
(1+\lambda_1)k_0^2, 
\label{eigenvalues02}
\end{equation}
where $k_0$ is a real number. Similarly, when $v^\mu = (0,{1\over {\sqrt 2}},{1\over {\sqrt 2}},0)$, the following differential operator takes the form
\begin{equation}
-{\partial^2_x} - {\partial^2_y}+\lambda_2\left(v\cdot\partial \right)^2=-(1-{\lambda_2\over 2})({\partial^2_x} + {\partial^2_y}),
\label{eigenvalues03}
\end{equation}
with plane waves eigenfunctions and the following eigenvalues 
\begin{equation}
(1-{\lambda_2\over 2})({k^2_x} + {k^2_y}), 
\label{eigenvalues04}
\end{equation}
where $k_x$ and $k_y$ are real numbers. The eigenfunctions of the operator $-\partial^2_z$ are standing waves, since these eigenfunctions are constrained to vanish at the two plates, $z=0$ and $z=a$, by the Dirichlet boundary conditions. Their eigenvalues are discrete and given by
\begin{equation}
\left({n\pi\over a}\right)^2, 
\label{eigenvalues05}
\end{equation}
where $n$ is a non-negative integer. Adding the quantities of Eqs. (\ref{eigenvalues02}), (\ref{eigenvalues04}), and (\ref{eigenvalues05}) to $m^2$, I find the eigenvalues of $D_E$ for Dirichlet boundary conditions
\begin{equation}
\left\{(1+\lambda_1)k_0^2+(1-{\lambda_2\over 2})(k_x^2+k_y^2)+\left({n\pi\over a}\right)^2+m^2\right\}.
\label{eigenvalues1}
\end{equation}
Also in the case of mixed boundary conditions the eigenfunctions of the operator $-\partial^2_z$ are standing waves, but these eigenfunctions are constrained by the mixed boundary conditions to vanish at one plate, $z=a$, and to have a vanishing derivative at the other plate, $z=0$. Therefore their discrete eigenvalues are
\begin{equation}
\left(n+{1\over 2}\right)^2\left({\pi\over a}\right)^2, 
\label{eigenvalues06}
\end{equation}
with $n$ a non-negative integer. The eigenvalues of the other terms in $D_E$ are the same as above, thus the
eigenvalues of $D_E$ for mixed boundary conditions are
\begin{equation}
\left\{(1+\lambda_1)k_0^2+(1-{\lambda_2\over 2})(k_x^2+k_y^2)+\left(n+{1\over 2}\right)^2\left({\pi\over a}\right)^2+m^2\right\}.
\label{eigenvalues2}
\end{equation}
\item Case {\bf \textcircled{2}}: $u$ is timelike and $v$ is $\parallel$ to $\bf B$.
I obtain the eigenvalues of $D_E$ for this situation using the same method I used for Case {\bf \textcircled{1}}, with some small differences in the algebra. I will not repeat the derivation here. The same is true for Case {\bf \textcircled{3}}. The
eigenvalues of $D_E$ for Dirichlet boundary conditions are
\begin{equation}
\left\{(1+\lambda_1)k_0^2+k_x^2+k_y^2+(1-\lambda_2)\left({n\pi\over a}\right)^2+m^2\right\},
\label{eigenvalues3}
\end{equation}
and the eigenvalues of $D_E$ for mixed boundary conditions are similar to those listed above, with $n$ replaced by $n+{1\over 2}$.

\item Case {\bf \textcircled{3}}: $u$ is $\perp$ to $\bf B$ and $v$ is $\parallel$ to $\bf B$.

Eigenvalues of $D_E$ for Dirichlet boundary conditions
\begin{equation}
\left\{k_0^2+(1-{\lambda_1\over 2})(k_x^2+k_y^2)+(1-\lambda_2)\left({n\pi\over a}\right)^2+m^2\right\},
\label{eigenvalues4}
\end{equation}

eigenvalues of $D_E$ for mixed boundary conditions are also obtained by replacing $n$ with $n+{1\over 2}$. 
\end{itemize}
In the eigenvalues listed above, $k_0, k_x, k_y \in \Re$ and $n=0,1,2,,\cdots$.

At this point I introduce the magnetic field. 
With  the presence of a constant magnetic field in the z-direction the spatial derivatives in the $x$ and $y$ direction, present in the Klein-Gordon operator of Eq. (\ref{KGoperator}), change and become
\begin{equation}
\vec{\nabla}_\perp\rightarrow (\vec{\nabla}-ie\vec{A})_\perp ,
\label{eigenvaluesnew1}
\end{equation}
where I use the notation $\vec{\nabla}_\perp=(\partial_x,\partial_y,0)$ and $\vec A$ is the electromagnetic vector potential which I write in terms of the electromagnetic field strength tensor as \begin{equation}
A^\mu={1\over 2}F^{\mu \nu}x_\nu .
\label{eigenvaluesnew2}
\end{equation}
For a constant magnetic field in the $z$-direction, the only non-vanishing components of $F^{\mu \nu}$ are 
\begin{equation}
F^{12}=-F^{21}=B .
\label{eigenvaluesnew3}
\end{equation}

The spectrum of the operator $-(\vec{\nabla}-ie\vec{A})_\perp^2$, where $\vec A$ is given by Eqs. (\ref{eigenvaluesnew2}) and (\ref{eigenvaluesnew3}), is well known from 
one-particle quantum mechanics and its eigenvalues are the Landau levels
\begin{equation}
2eB\left(\ell+{1\over 2}\right).
\label{eigenvaluesnew4}
\end{equation}
Therefore, the presence of the magnetic field modifies the eigenvalues of Eqs. (\ref{eigenvalues1} - \ref{eigenvalues4}) by changing $k^2_x+k^2_y$ into $(2\ell+1)eB$, where $\ell=0,1,2,\cdots$, labels the Landau levels. After making this modification to the eigenvalues, I write the generalized zeta function, with the magnetic field contribution, for the six aforementioned cases.
\begin{itemize}
\item Case {\bf \textcircled{1}} and Dirichlet boundary conditions
\begin{eqnarray}
\zeta(s)=&\mu^{2s}&{L^2eB\over 2\pi}(1+\lambda_1)\sum_{n=0}^\infty \sum_{\ell=0}^\infty\int^\infty_{-\infty} {dk_0\over 2\pi} \left[(1+\lambda_1)k_0^2+\left(1-{\lambda_2\over 2}\right)(2\ell+1)eB
\right.
\nonumber \\
&+&\left.
\left({n\pi\over a}\right)^2+m^2\right]^{-s},
\label{zeta1}
\end{eqnarray}
for mixed boundary conditions
\begin{eqnarray}
\zeta(s)=&\mu^{2s}&{L^2eB\over 2\pi}(1+\lambda_1)\sum_{n=0}^\infty \sum_{\ell=0}^\infty\int^\infty_{-\infty} {dk_0\over 2\pi} \left[(1+\lambda_1)k_0^2+\left(1-{\lambda_2\over 2}\right)(2\ell+1)eB
\right.
\nonumber \\
&+&\left.
\left(n+{1\over 2}\right)^2\left({\pi\over a}\right)^2+m^2\right]^{-s}.
\label{zeta1b}
\end{eqnarray}
In these equations the parameter $\mu$ with dimension of mass keeps $\zeta(s)$ dimensionless for all values of $s$ \cite{Hawking:1976ja}, and ${L^2eB\over 2\pi}$ takes into account the degeneracy of the Landau levels.
\item Case {\bf \textcircled{2}} and Dirichlet boundary conditions
\begin{eqnarray}
\zeta(s)=&\mu^{2s}&{L^2eB\over 2\pi}(1+\lambda_1)\sum_{n=0}^\infty \sum_{\ell=0}^\infty\int^\infty_{-\infty} {dk_0\over 2\pi} \left[(1+\lambda_1)k_0^2+(2\ell+1)eB\right.
\nonumber \\
&+&\left.
(1-\lambda_2)\left({n\pi\over a}\right)^2+m^2\right]^{-s},
\label{zeta2}
\end{eqnarray}
for mixed boundary conditions the zeta function is obtained by replacing $n$ with $n+{1\over 2}$ in the expression above.
\item Case {\bf \textcircled{3}} and Dirichlet boundary conditions
\begin{eqnarray}
\zeta(s)=&\mu^{2s}&{L^2eB\over 2\pi}\sum_{n=0}^\infty \sum_{\ell=0}^\infty\int^\infty_{-\infty} {dk_0\over 2\pi} \left[(1+\lambda_1)k_0^2+\left(1-{\lambda_1\over 2}\right)(2\ell+1)eB\right.
\nonumber \\
&+&\left.
(1-\lambda_2)\left({n\pi\over a}\right)^2+m^2\right]^{-s},
\label{zeta3}
\end{eqnarray}
and replacing $n$ with $n+{1\over 2}$ in the expression above yields the zeta function for mixed boundary conditions. 
\end{itemize}
In the next three sections, I will evaluate the zeta function and obtain the Casimir energy and pressure for Dirichlet and mixed boundary conditions for the three combinations of $u$ and $v$ listed above.
%%%%%%%%%%%%%%%%%%%%%%%%%%%%%%%%%%%%%%%%%%%%%%%%%%%%%%%%%%%%%%%%%%%%%
\section{Timelike anisotropy with spacelike anisotropy $\perp$ to $\bf B$}
\label{3}

This is what I called Case {\bf \textcircled{1}}, and I start by examining it for Dirichlet boundary conditions. 
I use the following identity
\begin{equation}
z^{-s}={1\over \Gamma(s)}\int_0^\infty dt \,t^{s-1}e^{-zt},
\label{zgamma}
\end{equation}
and rewrite Eq. (\ref{zeta1}) as
\begin{equation}
\zeta(s)={\mu^{2s}\over \Gamma(s)}{L^2eB\over 4\pi^2}\sqrt{1+\lambda_1}\sum_{n=0}^\infty \sum_{\ell=0}^\infty\int^\infty_{-\infty} \!\!\!\!dk_0 \int_0^\infty \!\!\!\! t^{s-1}e^{-\left[k_0^2+(1-{\lambda_2\over 2})(2\ell+1)eB+({n\pi\over a})^2+m^2\right]t} dt\, .
\label{zeta1_1}
\end{equation}
I integrate over $k_0$ and use this other identity,
\begin{equation}
\sum_{\ell=0}^\infty e^{-(1-{\lambda_2\over 2})(2\ell +1)eBt}={1\over 2 \sinh\left[(1-{\lambda_2\over 2})eBt\right]},
\label{sinh}
\end{equation}
to find
\begin{equation}
\zeta(s)={\mu^{2s}\over \Gamma(s)}{L^2eB\over 8\pi^{3\over 2}}\sqrt{1+\lambda_1}\sum_{n=1}^\infty \int_0^\infty \!\!\!\! t^{s-{3\over 2}}{e^{-\left[({n\pi\over a})^2+m^2\right]t} \over \sinh\left[(1-{\lambda_2\over 2})eBt\right] }dt\, ,
\label{zeta1_2}
\end{equation}
where I neglected the $n=0$ term because it does not depend on the plate distance $a$. 

When I investigate Case {\bf \textcircled{1}} for mixed boundary conditions, I take similar steps to those shown above and obtain the following zeta function
\begin{equation}
\zeta(s)={\mu^{2s}\over \Gamma(s)}{L^2eB\over 8\pi^{3\over 2}}\sqrt{1+\lambda_1}\sum_{n=1}^\infty \int_0^\infty \!\!\!\! t^{s-{3\over 2}}{e^{-\left[\left(n+{1\over 2}\right)^2({\pi\over a})^2+m^2\right]t} \over \sinh\left[(1-{\lambda_2\over 2})eBt\right] }dt\, .
\label{zeta1_1m}
\end{equation}

While these two zeta function cannot be evaluated in a closed form in the general case, they can be reduced to a simple analytic form in many limiting cases.

I begin by investigating the strong magnetic field limit, $\sqrt{eB}\gg a^{-1} , m$ under Dirichlet boundary conditions. In this approximation, I take
\begin{equation}
{1 \over \sinh\left[(1-{\lambda_2\over 2})eBt\right] }\simeq 2e^{-(1-{\lambda_2\over 2})eBt}
\label{sinh1}
\end{equation}
and do a Poisson resummation of Eq. (\ref{zeta1_2}) to obtain
\begin{equation}
\zeta(s)={\mu^{2s}\over \Gamma(s)}{eBL^2a\over 4\pi^{2}}{ \sqrt{1+\lambda_1}}\sum_{n=1}^\infty  \int_0^\infty \!\!\!\! t^{s-2}e^{-{n^2a^2\over t}}e^{-[(1-{\lambda_2\over 2})eB+m^2]t} dt,
\label{zeta1_3}
\end{equation}
where I neglected the term with $n=0$ since it is a uniform energy density term that does not contribute to the Casimir pressure if the magnetic field is present both inside and outside the plates. Next, I change the variable of integration in Eq. (\ref{zeta1_3}) to find
\begin{eqnarray}
\zeta(s)&=&{\mu^{2s}\over \Gamma(s)}{eBL^2a\over 4\pi^{2}}{ \sqrt{1+\lambda_1}}\sum_{n=1}^\infty  \left({an\over \sqrt{(1-{\lambda_2\over 2})eB+m^2}}\right)^{s-1}\nonumber \\
&\times&\int_0^\infty \!\!\!\! t^{s-2}e^{-na(t+1/t)\sqrt{(1-{\lambda_2\over 2})eB+m^2}} dt.
\label{zeta1_4}
\end{eqnarray}
Finally, I use the saddle point method to evaluate the integral and obtain
\begin{eqnarray}
\zeta(s)&=&{1\over \Gamma(s)}\left({\mu^{2}a\over \sqrt{(1-{\lambda_2\over 2})eB+m^2}}\right)^s{L^2\over 4\pi^{3/2}}{{\sqrt{1+\lambda_1}}eB}\nonumber \\
&\times&
{[(1-{\lambda_2\over 2})eB+m^2]^{1\over 4}\over \sqrt{a}} {e^{-2a\sqrt{(1-{\lambda_2\over 2})eB+m^2}}},
\label{zeta1_5}
\end{eqnarray}
where I neglected terms with $n>1$ because they are exponentially suppressed when compared to the $n=1$ term.

When examining the strong magnetic field case and mixed boundary conditions I proceed with similar steps. The Poisson resummation produces an extra factor of $(-1)^n$ inside the summation of Eq. (\ref{zeta1_4}) with everything else staying the same. Therefore, the mixed boundary conditions zeta function in the strong magnetic field limit is
\begin{eqnarray}
\zeta(s)=&-&{1\over \Gamma(s)}\left({\mu^{2}a\over \sqrt{(1-{\lambda_2\over 2})eB+m^2}}\right)^s{L^2\over 4\pi^{3/2}}{{\sqrt{1+\lambda_1}}eB}\nonumber \\
&\times&
{[(1-{\lambda_2\over 2})eB+m^2]^{1\over 4}\over \sqrt{a}} {e^{-2a\sqrt{(1-{\lambda_2\over 2})eB+m^2}}}.
\label{zeta1_5m}
\end{eqnarray}

Next I investigate the large mass limit, $m\gg a^{-1} , \sqrt{eB}$. In the case of Dirichlet boundary conditions, after a Poisson resummation of Eq. (\ref{zeta1_2}), I obtain
\begin{equation}
\zeta(s)={\mu^{2s}\over \Gamma(s)}{L^2a eB\over 8\pi^2}\sqrt{1+\lambda_1}\sum_{n=1}^\infty \int_0^\infty \!\!\!\! t^{s-2}{e^{-\left({n^2a^2\over t}+m^2t\right)} \over \sinh\left[(1-{\lambda_2\over 2})eBt\right] }dt\, ,
\label{zeta1_6}
\end{equation}
where, again, I neglected the $n=0$ term because it does not depend on the plate distance $a$. After a change of the integration variable, I find
\begin{equation}
\zeta(s)={\mu^{2s}\over \Gamma(s)}{L^2a eB\over 8\pi^2}\sqrt{1+\lambda_1}\sum_{n=1}^\infty\left({na\over m}\right)^{s-1} \int_0^\infty \!\!\!\! t^{s-2}{e^{-nam(t+1/t)} \over \sinh\left[(1-{\lambda_2\over 2}){eBna\over m}t\right] }dt\, ,
\label{zeta1_7}
\end{equation}
where all summation terms with $n>1$ are exponentially suppressed and can be neglected. Finally, I integrate using the saddle point method, to obtain
\begin{equation}
\zeta(s)={\mu^{2s}\over \Gamma(s)}\left({a\over m}\right)^{s}{L^2\over 8}\sqrt{{1+\lambda_1\over 1-\lambda_2}}\left({m\over \pi a}\right)^{3/2} {e^{-2am}}F\left[\left(1-{\lambda_2\over 2}\right)z\right]\, ,
\label{zeta1_8}
\end{equation}
where the dependence on the magnetic field is inside the function I called $F$ which I define as
\begin{equation}
F(z)={z\over \sinh (z)},
\label{F}
\end{equation}
with the dimensionless parameter $z$ defined as
\begin{equation}
z={eBa\over m},
\label{z}
\end{equation}
and where I used $1-{\lambda_2\over 2}\simeq \sqrt{{ 1-\lambda_2}}$, since $\lambda_2\ll1$.

The large mass limit calculation for mixed boundary conditions is very similar to the one outlined above that leads to Eq. (\ref{zeta1_8}). The one difference is that the starting point is the zeta function of Eq. (\ref{zeta1_1m}) and thus the Poisson resummation produces an extra factor of $(-1)^n$ into the summation. Following the other steps outlined above, I find
\begin{equation}
\zeta(s)=-{\mu^{2s}\over \Gamma(s)}\left({a\over m}\right)^{s}{L^2\over 8}\sqrt{{1+\lambda_1\over 1-\lambda_2}}\left({m\over \pi a}\right)^{3/2} {e^{-2am}}F\left[\left(1-{\lambda_2\over 2}\right)z\right]\, ,
\label{zeta1_8m}
\end{equation}
where the only difference with Eq. (\ref{zeta1_8}) is the overall negative sign.

Last, I will examine the small plate distance limit, $a^{-1}\ll m, \sqrt{eB}$. The details of this calculation, lengthier than the previous two, are shown in the Appendix. For the case of Dirichlet boundary conditions I find the following zeta function
\begin{eqnarray}
\zeta(s)&=&\left({\mu a\over \pi}\right)^{2s}{1\over \Gamma(s)}{L^2\over 8\pi^{3/2}}\sqrt{{1+\lambda_1\over 1-\lambda_2}}\left[{\pi^3\over a^3}\zeta_R(2s-3)\Gamma\left(s-{3\over 2}\right)-{\pi m^2\over a}\zeta_R(2s-1)\right.
\nonumber \\
&\times&\left.\Gamma\left(s-{1\over 2}\right)-{e^2B^2a\over 6\pi}(1-\lambda_2)\zeta_R(2s+1)\Gamma\left(s+{1\over 2}\right)
\right]\, ,
\label{zeta1_9}
\end{eqnarray}
where $\zeta_R(s)$ is the Riemann zeta function and where I used, as I did above, $\lambda_2\ll 1$. 

When considering mixed boundary conditions and the small plate distance limit, I obtain
\begin{eqnarray}
\zeta(s)&=&\left({\mu a\over \pi}\right)^{2s}{1\over \Gamma(s)}{L^2\over 8\pi^{3/2}}\sqrt{{1+\lambda_1\over 1-\lambda_2}}\left[{\pi^3\over a^3}\zeta_H\left(2s-3, {1\over 2}\right)\Gamma\left(s-{3\over 2}\right)-{\pi m^2\over a}\zeta_H\left(2s-1, {1\over 2}\right)\right.
\nonumber \\
&\times&\left.\Gamma\left(s-{1\over 2}\right)-{e^2B^2a\over 6\pi}(1-\lambda_2)\zeta_H\left(2s+1, {1\over 2}\right)\Gamma\left(s+{1\over 2}\right)
\right]\, ,
\label{zeta1_9m}
\end{eqnarray}
where $\zeta_H(s,x)$ is the Hurwitz zeta function defined as
\begin{equation}
\zeta_H(s,x) = \sum_{n=0}^\infty (n+x)^{-s}.
\label{Hurwitz}
\end{equation}
%%%%%%%%%%%%%%%%%%%%%%%%%%%%%%%%%%%%%%%%%%%%%%%%%%%%%%%%%%%%%%%%%%%%%
\subsection{Casimir energy and pressure for timelike anisotropy with spacelike anisotropy $\perp$ to $\bf B$}
\label{3_1}
The Casimir energy, $E_C$, of a complex scalar field confined between two parallel plates is related to the derivative of its zeta function by \cite{Hawking:1976ja} 
\begin{equation}
E_C = -\zeta'(0).
\label{E_C}
\end{equation}

To calculate the Casimir energy in the case of strong magnetic field and Dirichlet boundary conditions, I need to evaluate the derivative of the zeta function of Eq. (\ref{zeta1_5}) at $s=0$. This is easily done using the following, valid for $s\ll 1$,
\begin{equation}
{A^s\over \Gamma(s)}\simeq s+{\cal O}(s^2), 
\label{id4}
\end{equation}
where $A$ is any constant independent of $s$, to obtain
\begin{equation}
E_C =-{L^2eB\over 4\pi^{3/2}}{\sqrt{1+\lambda_1}}{[\sqrt{1-\lambda_2} eB+m^2]^{1\over 4}\over \sqrt{a}} {e^{-2a\sqrt{\sqrt{1-\lambda_2} eB+m^2}}},
\label{E_C1}
\end{equation}
which agrees with Ref. \cite{Erdas:2020ilo} once I take $\lambda_2=0$. Notice that the exponentially suppressed factor is the dominant term of the Casimir energy. 

The Casimir pressure is defined as
\begin{equation}
P_C=-{1\over L^2}{\partial E_C\over \partial a},
\label{P_C1}
\end{equation}
so, in the strong magnetic field limit, is
\begin{eqnarray}
P_C=&-&{eB\over 2\pi^{3/2}}{(\sqrt{1-\lambda_2}eB+m^2)^{3\over 4}\over {\sqrt a}} {\sqrt{1+\lambda_1}}e^{-2a\sqrt{\sqrt{1-\lambda_2}eB+m^2}}
\nonumber \\
&\times&\left(1+{1\over 4a\sqrt{\sqrt{1-\lambda_2}eB+m^2}}\right),
\label{P_C2}
\end{eqnarray}
where the dominant term is still the exponentially suppressed factor.

When I calculate $E_C$ and $P_C$ for strong magnetic field and mixed boundary conditions, the only change I need to make with respect to Eqs. (\ref{E_C1}) and (\ref{P_C2}) is to multiply by an extra factor of negative one, as I explain in the previous section, and therefore Casimir energy and pressure are
\begin{equation}
E_C ={L^2eB\over 4\pi^{3/2}}{\sqrt{1+\lambda_1}}{[\sqrt{1-\lambda_2} eB+m^2]^{1\over 4}\over \sqrt{a}} {e^{-2a\sqrt{\sqrt{1-\lambda_2} eB+m^2}}},
\label{EC_1m}
\end{equation}
and
\begin{eqnarray}
P_C&=&{eB\over 2\pi^{3/2}}{(\sqrt{1-\lambda_2}eB+m^2)^{3\over 4}\over {\sqrt a}} {\sqrt{1+\lambda_1}}e^{-2a\sqrt{\sqrt{1-\lambda_2}eB+m^2}}\nonumber \\
&\times&\left(1+{1\over 4a\sqrt{\sqrt{1-\lambda_2}eB+m^2}}\right).
\label{P_C2m}
\end{eqnarray}
Notice that in the case of mixed boundary conditions the force on the plates is repulsive.

In the large mass limit and under Dirichlet boundary conditions at the plates, I use Eqs. (\ref{E_C}), (\ref{id4}), and (\ref{P_C1})to find
\begin{equation}
E_C=-{L^2\over 8}\sqrt{{1+\lambda_1\over 1-\lambda_2}}\left({m\over \pi a}\right)^{3/2} {e^{-2am}}F\left(z\sqrt{1-\lambda_2}\right)\, ,
\label{E_C3}
\end{equation}
and
\begin{eqnarray}
P_C=&-&{m\over 4}\sqrt{{1+\lambda_1\over 1-\lambda_2}}\left({m\over \pi a}\right)^{3/2} {e^{-2am}}F\left(z\sqrt{1-\lambda_2}\right)
\nonumber \\
&\times&\left[1+{1\over 4 am}+{eB\sqrt{1-\lambda_2}\over 2m^2}\coth\left(z\sqrt{1-\lambda_2}\right)\right]\, ,
\label{P_C3}
\end{eqnarray}
where the dimensionless parameter $z$ and the function $F$ are defined in Eqs. (\ref{F}) and (\ref{z}). Notice that $z$ can be small or large, in the large mass limit, depending on the relative ratio of 
$\sqrt{eB}$ and $a^{-1}$, and therefore it is interesting to see a simpler approximate expression of the Casimir pressure in both cases. When $z\gg 1$ (i.e. ${eB\over m^2}\gg {1\over am}$) I can write $P_C$ of Eq. (\ref{P_C3}) as
\begin{equation}
P_C=-{m\over 2}\sqrt{{1+\lambda_1}}\left({m\over \pi^3 a}\right)^{1/2} {e^{-2am}} e^{-z\sqrt{1-\lambda_2}}eB\left(1+{eB\sqrt{1-\lambda_2}\over 2m^2}\right)\, .
\label{P_C4}
\end{equation}
When $z\ll 1$ (i.e. ${eB\over m^2}\ll {1\over am}$) the Casimir pressure of Eq. (\ref{P_C3}) takes this form
\begin{equation}
P_C=-{m\over 4}\sqrt{{1+\lambda_1\over 1-\lambda_2}}\left({m\over \pi a}\right)^{3/2} {e^{-2am}}\left[1+{3\over 4 am}-(1-\lambda_2){z^2\over 6}\right]\, .
\label{P_C5}
\end{equation}

When I examine mixed boundary conditions at the plates in the large mass limit, I obtain results similar to those listed above in Eqs. (\ref{E_C3}) - (\ref{P_C5}), with the only difference being that the negative sign in front of those four expressions becomes a positive sign.

The calculation of the small plate distance Casimir energy for Dirichlet boundary conditions is shown in the Appendix, where I obtain
\begin{equation}
{E_C} = -{\pi^2\over 8}{L^2\over a^3}\sqrt{{1+\lambda_1\over 1-\lambda_2}}\left[{1\over 90}-{m^2a^2\over 6\pi^2}-(1-\lambda_2){e^2B^2a^4\over 6\pi^4}\left(\gamma_E+\ln\left[{\mu a\over 2\pi}\right]\right)\right],
\label{E_C4}
\end{equation}
where  $\gamma_E=0.57721\cdots$ is the Euler-Mascheroni constant, and where the parameter $\mu$ takes the value $\mu = \sqrt{eB+m^2}$, as explained in Ref. \cite{Erdas:2021xvv}. I evaluate the Casimir pressure using Eq. (\ref{P_C1}) and obtain
\begin{equation}
{P_C} = -{\pi^2\over 8a^4}\sqrt{{1+\lambda_1\over 1-\lambda_2}}\left[{1\over 30}-{m^2a^2\over 6\pi^2}+(1-\lambda_2){e^2B^2a^4\over 6\pi^4}\left(1+\gamma_E+\ln\left[{\mu a\over 2\pi}\right]\right)\right].
\label{P_C6}
\end{equation}
Notice how the presence of a magnetic field increases the plate attractive pressure by an amount proportional to the square of the magnetic field alone and independent of the plate distance.

I calculate $E_C$ for mixed boundary conditions and small plate distance in the Appendix and find
\begin{equation}
{E_C} = {\pi^2\over 8}{L^2\over a^3}\sqrt{{1+\lambda_1\over 1-\lambda_2}}\left[{7\over 720}-{m^2a^2\over 12\pi^2}+(1-\lambda_2){e^2B^2a^4\over 6\pi^4}\left(\gamma_E+\ln\left[{2\mu a\over \pi}\right]\right)\right],
\label{E_C5}
\end{equation}
with the Casimir pressure given by
\begin{equation}
{P_C} = {\pi^2\over 8a^4}\sqrt{{1+\lambda_1\over 1-\lambda_2}}\left[{7\over 240}-{m^2a^2\over 12\pi^2}-(1-\lambda_2){e^2B^2a^4\over 6\pi^4}\left(1+\gamma_E+\ln\left[{2\mu a\over \pi}\right]\right)\right],
\label{P_C7}
\end{equation}
where  $\mu = \sqrt{eB+m^2}$. The pressure is repulsive in the case of mixed boundary conditions, as it happens in the Casimir effect without Lorentz violation. The magnetic field produces an attractive pressure
proportional to the square of $B$ and independent of $a$. This attractive "magnetic" pressure goes against the larger repulsive leading term.
%%%%%%%%%%%%%%%%%%%%%%%%%%%%%%%%%%%%%%%%%%%%%%%%%%%%%%%%%%%%%%%%%%%%%
\section{Timelike anisotropy with spacelike anisotropy $\parallel$ to $\bf B$}
\label{4}
For Case {\bf \textcircled{2}}, timelike anisotropy with spacelike anisotropy parallel to the magnetic field, I write the zeta function for Dirichlet boundary conditions of Eq. (\ref{zeta2}) as
\begin{equation}
\zeta(s)={\mu^{2s}\over \Gamma(s)}{L^2eB\over 4\pi^2}\sqrt{1+\lambda_1}\sum_{n=0}^\infty \sum_{\ell=0}^\infty\int^\infty_{-\infty} \!\!\!\!dk_0 \int_0^\infty \!\!\!\! t^{s-1}e^{-\left[k_0^2+(2\ell+1)eB+(1-{\lambda_2})({n\pi\over a})^2+m^2\right]t} dt\, .
\label{zeta2_1}
\end{equation}
When this zeta function is compared with the zeta function for timelike anisotropy with spacelike anisotropy $\perp$ to $\bf B$ of Eq, (\ref{zeta1_1}), it is evident that (\ref{zeta2_1}) can be obtained by taking (\ref{zeta1_1}) and, a) replacing $eB$ with $eB\over {1-{\lambda_2\over 2}}$ in it, b) multiplying it by an overall factor of $(1-{\lambda_2\over 2})$ and, c) replacing $a$ with $a\over \sqrt{1-{\lambda_2}}$ in it. Therefore $E_C$ for timelike anisotropy with spacelike anisotropy parallel to the magnetic field is obtained, in each of the three asymptotic limits  I consider in Sec. \ref{3}, by taking Eqs. (\ref{E_C1}), (\ref{E_C3}), (\ref{E_C4}) and making the same replacements. Once I do that, I obtain
\begin{equation}
E_C =-{L^2eB\over 4\pi^{3/2}}{\sqrt{1+\lambda_1}}\left[{eB+m^2\over a^2(1+\lambda_2)}\right]^{1/4} {e^{-2a\sqrt{1+\lambda_2}\sqrt{ eB+m^2}}},
\label{E_C2_1}
\end{equation}
and
\begin{equation}
P_C=-{eB\over 2\pi^{3/2}}{\sqrt{1+\lambda_1}}{(eB+m^2)^{3\over 4}\over {\sqrt {a\sqrt{1+\lambda_2}}}} e^{-2a\sqrt{1+\lambda_2}\sqrt{eB+m^2}}\left(1+{1\over 4a\sqrt{1+\lambda_2}\sqrt{eB+m^2}}\right),
\label{P_C2_1}
\end{equation}
both valid for strong magnetic field. Notice that the presence of timelike anisotropy increases the Casimir pressure, while the presence of spacelike anisotropy parallel to $\bf B$ decreases the Casimir pressure.

In the limit of large mass, after applying the aforementioned modifications to the appropriate equations, I find
\begin{equation}
E_C=-{L^2\over 8}\sqrt{1+\lambda_1}\left({m\over \pi a\sqrt{1+\lambda_2}}\right)^{3/2} {e^{-2a\sqrt{1+\lambda_2}m}}F\left(z\sqrt{1+\lambda_2}\right)\, ,
\label{E_C2_2}
\end{equation}
and
\begin{eqnarray}
P_C=&-&{m\over 4}\sqrt{1+\lambda_1}\left({m\over \pi a\sqrt{1+\lambda_2}}\right)^{3/2} {e^{-2a\sqrt{1+\lambda_2}m}}F\left(z\sqrt{1+\lambda_2}\right)
\nonumber \\
&\times&\left[1+{1\over 4 a\sqrt{1+\lambda_2}m}+{eB\over 2m^2}\coth\left(z\sqrt{1+\lambda_2}\right)\right]\, ,
\label{P_C2_2}
\end{eqnarray}
where, again, the presence of timelike anisotropy increases the pressure and the presence of spacelike anisotropy parallel to $\bf B$ decreases it. As I did in the previous section, I will write down $P_C$ for $z\gg 1$
\begin{equation}
P_C=-{m\over 2}\sqrt{{1+\lambda_1}}\left({m\over \pi^3 a\sqrt{1+\lambda_2}}\right)^{1/2} {e^{-2a\sqrt{1+\lambda_2}m}} e^{-z\sqrt{1+\lambda_2}}eB\left(1+{eB\over 2m^2}\right)\, ,
\label{P_C2_3}
\end{equation}
and for $z\ll 1$
\begin{equation}
P_C=-{m\over 4}\sqrt{1+\lambda_1}\left({m\over \pi a\sqrt{1+\lambda_2}}\right)^{3/2} {e^{-2a\sqrt{1+\lambda_2}m}}\left[1+{3\over 4 a\sqrt{1+\lambda_2}m}-(1+\lambda_2){z^2\over 6}\right]\, .
\label{P_C2_4}
\end{equation}

In the small plate distance approximation, I obtain
\begin{eqnarray}
{E_C} = &-&{\pi^2\over 8}{L^2\over a^3}\sqrt{{1+\lambda_1}\over {(1+\lambda_2)^3}}\left[{1\over 90}-{m^2a^2\over 6\pi^2}(1+\lambda_2)
\right.
\nonumber \\
&-&\left.{e^2B^2a^4\over 6\pi^4}(1+\lambda_2)^2\left(\gamma_E+\ln\left[{\mu a\sqrt{1+\lambda_2}\over 2\pi}\right]\right)\right],
\label{E_C2_3}
\end{eqnarray}
and
\begin{eqnarray}
{P_C} = &-&{\pi^2\over 8a^4}{\sqrt{1+\lambda_1}\over (1+\lambda_2)^2}\left[{1\over 30}-{m^2a^2\over 6\pi^2}(1+\lambda_2)
\right.
\nonumber \\
&+&\left.{e^2B^2a^4\over 6\pi^4}(1+\lambda_2)^2\left(1+\gamma_E+\ln\left[{\mu a\sqrt{1+\lambda_2}\over 2\pi}\right]\right)\right],
\label{P_C2_5}
\end{eqnarray}
where, once more, the presence of timelike anisotropy increases the pressure while the presence of spacelike anisotropy tends to decrease it.

When I investigate mixed boundary conditions for Case {\bf \textcircled{2}}, the zeta function is the same as that of Eq. (\ref{zeta2_1}), with $n$ replaced by $n+{1\over 2}$. The same replacements I list above, applied to the Casimir energy equations for mixed boundary conditions of Sec. (\ref{3}), produce the mixed boundary conditions Casimir energy in the three approximations I consider. I find that, in the strong magnetic field approximation, the mixed boundary conditions Casimir energy and pressure are given by Eqs. (\ref{E_C2_1}) and (\ref{P_C2_1}) multiplied by $-1$. The mixed boundary conditions Casimir energy and pressure in the large mass limit are obtained by taking Eqs. (\ref{E_C2_2}), (\ref{P_C2_2}) and multiplying them by $-1$. The small distance approximations of $E_C$ and $P_C$ under mixed boundary conditions are obtained by taking Eqs. (\ref{E_C5}) and (\ref{P_C7}) and modifying them as I describe at the beginning of this section. Once I do that, I obtain
\begin{eqnarray}
{E_C} &=& {\pi^2\over 8}{L^2\over a^3}\sqrt{{1+\lambda_1}\over {(1+\lambda_2)^3}}\left[{7\over 720}-{m^2a^2\over 12\pi^2}(1+\lambda_2)
\right.
\nonumber \\
&+&\left.{e^2B^2a^4\over 6\pi^4}(1+\lambda_2)^2\left(\gamma_E+\ln\left[{2\mu a\sqrt{1+\lambda_2}\over \pi}\right]\right)\right],
\label{E_C2_4}
\end{eqnarray}
and
\begin{eqnarray}
{P_C} &=& {\pi^2\over 8a^4}{\sqrt{1+\lambda_1}\over (1+\lambda_2)^2}\left[{7\over 240}-{m^2a^2\over 12\pi^2}(1+\lambda_2)
\right.
\nonumber \\
&-&\left.{e^2B^2a^4\over 6\pi^4}(1+\lambda_2)^2\left(1+\gamma_E+\ln\left[{2\mu a\sqrt{1+\lambda_2}\over \pi}\right]\right)\right],
\label{P_C2_6}
\end{eqnarray}
notice that, also in this case, the  presence of timelike anisotropy increases the repulsive pressure while the presence of spacelike anisotropy goes against it.
%%%%%%%%%%%%%%%%%%%%%%%%%%%%%%%%%%%%%%%%%%%%%%%%%%%%%%%%%%%%%%%%%%%%%
\section{Spacelike anisotropy $\perp$ to $\bf B$ with spacelike anisotropy $\parallel$ to $\bf B$}
\label{5}
When investigating Case {\bf \textcircled{3}}, spacelike anisotropy perpendicular to $\bf B$ with spacelike anisotropy parallel to $\bf B$, I write the zeta function for Dirichlet boundary conditions of Eq. (\ref{zeta3}) as
\begin{equation}
\zeta(s)={\mu^{2s}\over \Gamma(s)}{L^2eB\over 4\pi^2}\sum_{n=0}^\infty \sum_{\ell=0}^\infty\int^\infty_{-\infty} \!\!\!\!dk_0 \int_0^\infty \!\!\!\! t^{s-1}e^{-\left[k_0^2+(1-{\lambda_1\over 2})(2\ell+1)eB+(1-{\lambda_2})({n\pi\over a})^2+m^2\right]t} dt\, ,
\label{zeta3_1}
\end{equation}
and, when I compare this zeta function to the zeta function for timelike anisotropy with spacelike anisotropy $\perp$ to $\bf B$ of Eq. (\ref{zeta1_1}), it is clear that (\ref{zeta3_1}) can be obtained from  
(\ref{zeta1_1}) by a) dividing it by $\sqrt{1+\lambda_1}$, b) by changing $\lambda_2$ into $\lambda_1$ in it, and c) by replacing $a$ with $a\sqrt{1+\lambda_2}$. Below are my results for $E_C$ and $P_C$ under the three approximations I examine.

Under the strong magnetic field approximation, I find
\begin{equation}
E_C =-{L^2eB\over 4\pi^{3/2}}{[\sqrt{1-\lambda_1} eB+m^2]^{1\over 4}\over \sqrt{a\sqrt{1+\lambda_2}}} {e^{-2a\sqrt{1+\lambda_2}\sqrt{\sqrt{1-\lambda_1} eB+m^2}}},
\label{E_C3_1}
\end{equation}
and
\begin{eqnarray}
P_C=&-&{eB\over 2\pi^{3/2}}{(\sqrt{1-\lambda_1}eB+m^2)^{3\over 4}\over {\sqrt{ a \sqrt{1+\lambda_2}}}} e^{-2a\sqrt{1+\lambda_2}\sqrt{\sqrt{1-\lambda_1}eB+m^2}}
\nonumber \\
&\times&\left(1+{1\over 4a\sqrt{1+\lambda_2}\sqrt{\sqrt{1-\lambda_1}eB+m^2}}\right),
\label{P_C3_1}
\end{eqnarray}
notice how the presence of spacelike anisotropy parallel to $\bf B$ weakens the attractive Casimir pressure, as it was observed in the previous section.

When I consider the large mass limit, I obtain
\begin{equation}
E_C=-{L^2\over 8}\sqrt{1+\lambda_1}\left({m\over \pi a\sqrt{1+\lambda_2}}\right)^{3/2} {e^{-2a\sqrt{1+\lambda_2}m}}F\left(z\sqrt{1-\lambda_1}\sqrt{1+\lambda_2}\right)\, ,
\label{E_C3_2}
\end{equation}
and
\begin{eqnarray}
P_C&=&-{m\over 4}\sqrt{1+\lambda_1}\left({m\over \pi a\sqrt{1+\lambda_2}}\right)^{3/2} {e^{-2a\sqrt{1+\lambda_2}m}}F\left(z\sqrt{1-\lambda_1}\sqrt{1+\lambda_2}\right)\nonumber \\
&&\times\left[1+{1\over 4 a\sqrt{1+\lambda_2}m}+{eB\sqrt{1-\lambda_1}\over 2m^2}\coth\left(z\sqrt{1-\lambda_1}\sqrt{1+\lambda_2}\right)\right]\, ,
\label{P_C3_2}
\end{eqnarray}
where the presence of spacelike anisotropy perpendicular to $\bf B$ increases the attractive pressure while the presence of spacelike anisotropy parallel to $\bf B$ decreases it.

Next, in the small plate distance approximation
\begin{eqnarray}
{E_C} = &-&{\pi^2\over 8}{L^2\over a^3}\sqrt{1+\lambda_1\over ({1+\lambda_2})^3}\left[{1\over 90}-{m^2a^2\over 6\pi^2}(1+\lambda_2)-{e^2B^2a^4\over 6\pi^4}(1-\lambda_1)(1+\lambda_2)^2\right.
\nonumber \\
&\times&\left.
\left(\gamma_E+\ln\left[{\mu a\sqrt{1+\lambda_2}\over 2\pi}\right]\right)\right],
\label{E_C3_3}
\end{eqnarray}
\begin{eqnarray}
{P_C} = &-&{\pi^2\over 8a^4}{\sqrt{1+\lambda_1}\over ({1+\lambda_2})^2}\left[{1\over 30}-{m^2a^2\over 6\pi^2}(1+\lambda_2)+{e^2B^2a^4\over 6\pi^4}(1-\lambda_1)(1+\lambda_2)^2
\right.
\nonumber \\
&\times&\left.
\left(1+\gamma_E+\ln\left[{\mu a\sqrt{1+\lambda_2}\over 2\pi}\right]\right)\right],
\label{P_C3_3}
\end{eqnarray}
and, here too, the presence of spacelike anisotropy perpendicular to $\bf B$ augments the attractive force, while the presence of spacelike anisotropy parallel to $\bf B$ weakens it.

Finally, I examine mixed boundary conditions for Case {\bf \textcircled{3}}. In this case, the zeta function is the same as that of Eq. (\ref{zeta3_1}), with $n$ replaced by $n+{1\over 2}$. When comparing it to the zeta function for mixed boundary conditions of Case {\bf \textcircled{1}}, the same modifications listed at the beginning of this section need to be applied to the Case {\bf \textcircled{1}} zeta function to obtain the Case {\bf \textcircled{3}} one. In the strong magnetic field approximation $E_C$ and $P_C$ are obtained by multiplying by $-1$ the expressions of Eqs. (\ref{E_C3_1}) and (\ref{P_C3_1}). $E_C$ and $P_C$ in the large mass approximation are obtained in the same way, but using Eqs. (\ref{E_C3_2}) and (\ref{P_C3_2}). The Casimir energy and pressure in the small plate distance approximation are:
\begin{eqnarray}
{E_C} &=& {\pi^2\over 8}{L^2\over a^3}\sqrt{1+\lambda_1\over ({1+\lambda_2})^3}\left[{7\over 720}-{m^2a^2\over 12\pi^2}(1+\lambda_2)+{e^2B^2a^4\over 6\pi^4}(1-\lambda_1)(1+\lambda_2)^2
\right.
\nonumber \\
&\times&\left.
\left(\gamma_E+\ln\left[{2\mu a\sqrt{1+\lambda_2}\over \pi}\right]\right)\right],
\label{E_C3_4}
\end{eqnarray}
\begin{eqnarray}
{P_C} &=& {\pi^2\over 8a^4}{\sqrt{1+\lambda_1}\over ({1+\lambda_2})^2}\left[{7\over 240}-{m^2a^2\over 12\pi^2}(1+\lambda_2)-{e^2B^2a^4\over 6\pi^4}(1-\lambda_1)(1+\lambda_2)^2
\right.
\nonumber \\
&\times&\left.
\left(1+\gamma_E+\ln\left[{2\mu a\sqrt{1+\lambda_2}\over \pi}\right]\right)\right],
\label{P_C3_4}
\end{eqnarray}
where the presence of spacelike anisotropy perpendicular to $\bf B$ increases the repulsive pressure and the presence of spacelike anisotropy parallel to $\bf B$ weakens the repulsion.
%%%%%%%%%%%%%%%%%%%%%%%%%%%%%%%%%%%%%%%%%%%%%%%%%%%%%
\section{Discussion and conclusions}
\label{6}

In this paper, using the generalized zeta function technique, I studied the Casimir effect due to a scalar field between parallel plates that breaks Lorentz symmetry in two orthogonal directions when in the presence of a magnetic field. This complex scalar field has mass and charge
and satisfies a modified Klein-Gordon equation that breaks the Lorentz symmetry in an aether-like and CPT even manner, through a direct coupling of the field derivatives to two orthogonal unit vectors $u^\mu$ and $v^\mu$ with fixed spacetime directions. The strength of the coupling of the field derivatives to $u^\mu$ and $v^\mu$ is determined by the two dimensionless parameters $\lambda_1$ and $\lambda_2$, respectively. 

I investigated all possible combinations of directions of $u^\mu$ and $v^\mu$ and the case of the scalar field satisfying either Dirichlet or mixed (Dirichlet-Neumann) boundary conditions on the plates. The case of Neumann boundary conditions is completely straightforward since all results obtained for Dirichlet boundary conditions apply verbatim to the case of Neumann. There are three possible combinations of $u^\mu$ and $v^\mu$ orientations, I called them Case {\bf \textcircled{1}} (timelike anisotropy with spacelike anisotropy perpendicular to $\bf B$), Case {\bf \textcircled{2}} (timelike anisotropy with spacelike anisotropy parallel to $\bf B$), Case {\bf \textcircled{3}} (spacelike anisotropy perpendicular to $\bf B$ with spacelike anisotropy parallel to $\bf B$). I examined Case {\bf \textcircled{1}} in Sec. \ref{3}, Case {\bf \textcircled{2}} in Sec. \ref{4}, and Case {\bf \textcircled{3}} in Sec. \ref{5}. For each of these three cases, I obtained simple analytic expressions of the Casimir energy, Eqs. (\ref{E_C1}), (\ref{E_C2_1}), (\ref{E_C3_1}), and pressure (\ref{P_C2}), (\ref{P_C2_1}), (\ref{P_C3_1}) in the strong magnetic field approximation, large mass approximation, Eqs. (\ref{E_C3}), (\ref{E_C2_2}), (\ref{E_C3_2}), and (\ref{P_C3}), (\ref{P_C2_2}), (\ref{P_C3_2}), and small plate distance approximation Eqs. (\ref{E_C4}), (\ref{E_C2_3}), (\ref{E_C3_3}), and (\ref{P_C6}), (\ref{P_C2_5}), (\ref{P_C3_3}), for Dirichlet boundary conditions. When investigating mixed boundary conditions my results, in several cases, were the same results I obtained for Dirichlet boundary conditions, but with opposite sign. In those cases I just stated that without writing equations. In the small plate distance approximation however, for all three cases, the results I obtained for mixed boundary conditions were different from those I obtained for Dirichlet boundary conditions by more than an overall sign. $E_C$ and $P_C$ for mixed boundary conditions and all three possible orientations of $u^\mu$ and $v^\mu$ are shown in Eqs. (\ref{E_C5}), (\ref{E_C2_4}), (\ref{E_C3_4}), and (\ref{P_C7}), (\ref{P_C2_6}), (\ref{P_C3_4}), respectively.

My paper shows that, in the presence of a magnetic field and double Lorentz asymmetry, the Casimir pressure is attractive for Dirichlet (and Neumann) boundary conditions, and repulsive for mixed boundary conditions, as it happens in the standard Casimir effect without Lorentz asymmetry and magnetic field. It shows also a strong dependence of $E_C$ and $P_C$ on $\lambda_1$ and $\lambda_2$, the two dimensionless quantities that parametrize the Lorentz asymmetry. This strong dependence was also observed in Refs. \cite{deMello:2022tuv,Erdas:2020ilo,Erdas:2021xvv,Cruz:2017kfo}. My results fully agree with Ref. \cite{deMello:2022tuv}, when the magnetic field is switched off. I find that that the Casimir energy and pressure are exponentially suppressed in the strong magnetic field approximation in all three cases considered. The same is true for $E_C$ and $P_C$ in the large mass limit. In the small plate distance approximation, I find that the correction due to the magnetic field increases the attractive pressure in the case of Dirichlet boundary conditions, while it decreases the repulsion in the case of mixed boundary conditions, and this happens for all three combinations of orientation of $u^\mu$ and $v^\mu$. This correction depends only on the magnetic field and on the dimensionless parameters $\lambda_1$ and $\lambda_2$, not on the plate distance, and is strongest in Case {\bf \textcircled{2}}, timelike anisotropy with spacelike anisotropy parallel to $\bf B$.

Finally, since pressure is the physically measurable quantity, it is important to compare the pressure results obtained for strong magnetic field for Cases {\bf \textcircled{1}}, {\bf \textcircled{2}}, and {\bf \textcircled{3}}, the pressure for large mass for the three cases and the pressure for small plate distance for those three cases. I will start by focusing on Dirichlet boundary conditions and will also briefly discuss mixed boundary conditions, since the mixed boundary conditions results are similar to the Dirichlet case.

Comparing my results for the Casimir pressure under strong magnetic field and Dirichlet boundary conditions of Eqs. (\ref{P_C2}), (\ref{P_C2_1}), (\ref{P_C3_1}), we see how the presence of timelike anisotropy always produces a pressure increase relative to the standard Casimir effect in symmetric spacetime.
It is also interesting to point out that, when comparing those three equations, the presence of spacelike anisotropy perpendicular to the magnetic field always produces a stronger pressure than the standard Casimir effect in symmetric spacetime, while the presence of spacelike anisotropy parallel to the magnetic field
always produces a weaker pressure than in the standard Casimir effect. This is also true for mixed boundary conditions and strong magnetic field, the only difference being that the pressure is repulsive, not attractive.

When examining my results for the pressure in the large mass limit and under Dirichlet boundary conditions, Eqs. (\ref{P_C3}), (\ref{P_C2_2}), (\ref{P_C3_2}), one observes the same dependence of the pressure on the three types of anisotropy that I described above. However, the presence of spacelike anisotropy parallel to the magnetic field weakens the pressure more than the other two types of anisotropy increase it, since it augments an exponential suppression factor while timelike anisotropy and spacelike anisotropy simply produce an increasing multiplication factor. Again, this is true for mixed boundary conditions as well.

The results I obtained for the pressure under small plate distance approximation and Dirichlet boundary conditions, shown in Eqs. (\ref{P_C6}), (\ref{P_C2_5}), (\ref{P_C3_3}), confirm, once more, that timelike anisotropy and spacelike anisotropy perpendicular to $\bf B$ tend to increase the pressure when compared to that obtained for a symmetric spacetime, while spacelike anisotropy parallel to $\bf B$ tends to decrease it. The same is true for mixed boundary conditions.

Summarizing, this work shows that the presence of timelike anisotropy always produces a stronger Casimir pressure, under all approximations considered, when compared to the standard Casimir effect in Lorentz symmetric spacetime. This is true for the attractive pressure present when the scalar field satisfies Dirichlet boundary conditions at the plates, and for the repulsive pressure on the plates present when mixed boundary conditions are considered. The presence of spacelike anisotropy perpendicular to $\bf B$ also produces a stronger Casimir pressure than the pressure of the standard effect in symmetric spacetime. This happens under all conditions considered, and for both Dirichlet and mixed boundary conditions. I also find that the presence of spacelike anisotropy parallel to the magnetic field weakens the Casimir pressure when comparing it to the pressure in Lorentz symmetric spacetime, under all conditions considered and for both Dirichlet and mixed boundary conditions.
%%%%%%%%%%%%%%%%%%%%%%%%%%%%%%%%%%%%%%%%%%%%%%%%%%%%%%%%%%%%%%%%%%%%%
%%%%%%%%%%%%%%%%%%%%%%%%%%%%%%%%%%%%%%%%%%%%%%%%%%%%%
\section{Appendix}
\label{7}
In this appendix I will calculate the approximate zeta function and Casimir energy for timelike anisotropy with spacelike anisotropy perpendicular to $\bf B$ in the limit of small plate distance, $a^{-1}\ll m, \sqrt{eB}$. I start by examining the case of Dirichlet boundary conditions at the plates. In the small plate distance limit, I can make the following approximations
\begin{equation}
e^{-m^2t} \simeq 1-m^2t,
\label{A1}
\end{equation}
\begin{equation}
{eB\over \sinh\left[\left(1-{\lambda_2 \over 2}\right)eBt\right]}
\simeq {1\over \left(1-{\lambda_2 \over 2}\right)t} - \left(1-{\lambda_2 \over 2}\right){e^2B^2t\over 6},
\label{A2}
\end{equation}
and, when I use them inside Eq. (\ref{zeta1_2}), I find
\begin{equation}
\zeta(s)\simeq{\mu^{2s}\over \Gamma(s)}{L^2eB\over 8\pi^{3\over 2}}\sqrt{{1+\lambda_1\over 1-\lambda_2}}\sum_{n=1}^\infty \int_0^\infty \!\!\!\! t^{s-{5\over 2}}{e^{-({n\pi\over a})^2t} }\left[1-m^2t- \left(1-{\lambda_2}\right){e^2B^2t^2\over 6}\right]dt\, ,
\label{A3}
\end{equation}
where I neglect a smaller term proportional to $(meB)^2$. I do the integration and obtain the approximate zeta function
\begin{eqnarray}
\zeta(s)&=&\left({\mu a\over \pi}\right)^{2s}{1\over \Gamma(s)}{L^2\over 8\pi^{3/2}}\sqrt{{1+\lambda_1\over 1-\lambda_2}}\left[{\pi^3\over a^3}\zeta_R(2s-3)\Gamma\left(s-{3\over 2}\right)-{\pi m^2\over a}\zeta_R(2s-1)\right.
\nonumber \\
&\times&\left.\Gamma\left(s-{1\over 2}\right)-{e^2B^2a\over 6\pi}(1-\lambda_2)\zeta_R(2s+1)\Gamma\left(s+{1\over 2}\right)
\right]\, .
\label{A4}
\end{eqnarray}

When I consider mixed boundary conditions, $n$ in Eq. (\ref{A3}) is replaced by $n+{1\over 2}$ while the rest stays the same and I find the following approximate zeta function
\begin{eqnarray}
\zeta(s)&=&\left({\mu a\over \pi}\right)^{2s}{1\over \Gamma(s)}{L^2\over 8\pi^{3/2}}\sqrt{{1+\lambda_1\over 1-\lambda_2}}\left[{\pi^3\over a^3}\zeta_H\left(2s-3, {1\over 2}\right)\Gamma\left(s-{3\over 2}\right)-{\pi m^2\over a}\zeta_H\left(2s-1, {1\over 2}\right)\right.
\nonumber \\
&\times&\left.\Gamma\left(s-{1\over 2}\right)-{e^2B^2a\over 6\pi}(1-\lambda_2)\zeta_H\left(2s+1, {1\over 2}\right)\Gamma\left(s+{1\over 2}\right)
\right]\, .
\label{A5}
\end{eqnarray}

To evaluate the Casimir energy in the small plate distance for Dirichlet boundary conditions, I use the zeta function of Eq. (\ref{A4}) and, according to Eq. (\ref{E_C}), calculate its derivative at $s=0$. To take that derivative I use the following approximations, valid for $s\ll 1$
\begin{equation}
A^{2s}{\Gamma(s-{\textstyle\frac{3}{2}})\over\Gamma(s)}\zeta_R(2s-3)\simeq {\sqrt{\pi}\over 90}s +{\cal O}(s^2),
\label{id5}
\end{equation}
\begin{equation}
A^{2s}{\Gamma(s-{\textstyle\frac{1}{2}})\over\Gamma(s)}\zeta_R(2s-1)\simeq {\sqrt{\pi}\over 6}s +{\cal O}(s^2),
\label{id6}
\end{equation}
\begin{equation}
A^{2s}{\Gamma(s+{\textstyle\frac{1}{2}})\over\Gamma(s)}\zeta_R(2s+1)\simeq {\sqrt{\pi}\over 2}+{\sqrt{\pi}}\left[\gamma_E+\ln\left({A\over 2}\right)\right]s +{\cal O}(s^2),
\label{id7}
\end{equation}
where $A$ is a constant independent of $s$ and $\gamma_E=0.57721\cdots$ is the Euler-Mascheroni constant. I obtain
\begin{equation}
{E_C} = -{\pi^2\over 8}{L^2\over a^3}\sqrt{{1+\lambda_1\over 1-\lambda_2}}\left[{1\over 90}-{m^2a^2\over 6\pi^2}-(1-\lambda_2){e^2B^2a^4\over 6\pi^4}\left(\gamma_E+\ln\left[{\mu a\over 2\pi}\right]\right)\right],
\label{A6}
\end{equation}
where the parameter $\mu$ takes the value $\mu = \sqrt{eB+m^2}$, as explained in Ref. \cite{Erdas:2021xvv}. 

Now I consider mixed boundary conditions and evaluate $E_C$ in the limit of small plate distance. I use Eq. (\ref{A5}) and the following approximations
\begin{equation}
A^{2s}{\Gamma(s-{\textstyle\frac{3}{2}})\over\Gamma(s)}\zeta_H\left(2s-3,{1\over 2}\right)\simeq -{7\sqrt{\pi}\over 720}s +{\cal O}(s^2),
\label{id8}
\end{equation}
\begin{equation}
A^{2s}{\Gamma(s-{\textstyle\frac{1}{2}})\over\Gamma(s)}\zeta_H\left(2s-1,{1\over 2}\right)\simeq-{\sqrt{\pi}\over 12}s +{\cal O}(s^2),
\label{id9}
\end{equation}
\begin{equation}
A^{2s}{\Gamma(s+{\textstyle\frac{1}{2}})\over\Gamma(s)}\zeta_H\left(2s+1,{1\over 2}\right)\simeq {\sqrt{\pi}\over 2}+{\sqrt{\pi}}\left[\gamma_E+\ln\left({2A}\right)\right]s +{\cal O}(s^2),
\label{id10}
\end{equation}
to obtain
\begin{equation}
{E_C} = {\pi^2\over 8}{L^2\over a^3}\sqrt{{1+\lambda_1\over 1-\lambda_2}}\left[{7\over 720}-{m^2a^2\over 12\pi^2}+(1-\lambda_2){e^2B^2a^4\over 6\pi^4}\left(\gamma_E+\ln\left[{2\mu a\over \pi}\right]\right)\right],
\label{A7}
\end{equation}
with  $\mu = \sqrt{eB+m^2}$.
%%%%%%%%%%%%%%%%%%%%%%%%%%%%%%%%%%%%%%%%%%%%%%%%%%%%%

%%%%%%%%%%%%%%%%%%%%%%%%%%%%%%%%%%%%%%%%%%%%%%%%%%%%%%%%%%%%%%%%%%%
\end{document}